\documentclass[preprint,aps,nofootinbib]{revtex4}
\usepackage{}
\usepackage{graphicx}
\usepackage{amsmath}
\usepackage{amsfonts}
\usepackage{amssymb}
\usepackage{color}%
\usepackage{dcolumn}
\usepackage{indentfirst}
\providecommand{\U}[1]{\protect\rule{.1in}{.1in}}

\definecolor{lightgray}{rgb}{.7,.7,.7}

\definecolor{red}{rgb}{1,0,0}

\definecolor{blue}{rgb}{0,0,1}

\definecolor{purple}{rgb}{0.6,0.1,0.7}

\newcommand{\f}{\begin{equation}}
\newcommand{\ff}{\end{equation}}
\newcommand{\fa}{\begin{eqnarray}}
\newcommand{\ffa}{\end{eqnarray}}

\begin{document}
\title{Holographic Complexity Growth Rate in a dual FLRW Universe}
\author{Wen-Jian Pan $^{1}$}
\email{wjpan_zhgkxy@163.com}
\author{Yu-lin Li $^{1}$}
\author{Min Song $^{1}$}
\author{Wen-bin Xie $^{1}$}
\author{Shuang Zhang $^{1}$}
\affiliation{$^1$ College of Science,BeiBu Gulf University,Qinzhou,535011,People's Republic of China}

\begin{abstract}
 In this paper, taking the large $R$ limit and using the complexity-volume duality,
 we investigate the holographic complexity growth rate of a field state defined on the universe
 located at an asymptotical AdS boundary in Gauss-Bonnet gravity and massive gravity,
 respectively. For the Gauss-Bonnet gravity case, its growth behavior of the state
 mainly presents three kinds of contributions: one, as a finite term viewed as an interaction term, comes from
 a conserved charge, the second one is from the spatial volume of the universe
  and the third one relates the curvature of the horizon in the AdS Gauss-Bonnet black hole,
  where the Gauss-Bonnet effect plays a vital role on such growth rate. For massive gravity case,
  except the first divergent term still obeying the growth rate of the spatial volume of
the Universe, its results reveal the more interesting novel phenomenons: beside
the conserved charge $E$, the graviton mass term also provides its effect to the finite term
; and the third divergent term is determined by the spatial curvature of its horizon
$k$ and graviton mass effect; furthermore, the graviton mass effect can be completely responsible for the second divergent
term as a new additional term saturating an area law.

\end{abstract}
\maketitle

\section{Introduction}
Anti-de Sitter/Conformal field theory (AdS/CFT) correspondence
\cite{Maldacena:1997re,Gubser:1998bc,Witten:1998qj,Aharony:1999ti} is currently established
as a valuable prescription to approach the understanding of the quantum gravity.
Remarkably, the fascinating idea is that, by mapping physical degrees a strong coupled quantum system to
dual gravity theory in a higher dimensional bulk space, a difficult problem is usually transformed into
a tractable one. There has been extensively investigated
in the modern theoretical physics over the last decades years.
Specially, when it comes to the quantum information theory in the context
of AdS/CFT correspondence in recent years, one famous topic in this direction is
the holographic entanglement entropy proposed by Ryu and Takayanaki\cite{Ryu:2006bv},
which asserts that its quantum entanglement entropy of a conformal field theory
in the subregion on the boundary can be described equivalently by the minimal area
of a bulk codimension two surface anchored at the boundaries of the subregion.
This indeed reveals the dual relationship between quantum information
theory defined on boundary and gravity in bulk.

In the holographic context of a thermo-field double state
(TFD state)on the boundary theory being dual to a eternal black hole \cite{Maldacena:2001kr},
it has told us that the entanglement entropy can not capture
all the information for the full time evolution of an AdS wormhole\cite{Hartman:2013qma}.
As a result, an another refined information quantity, namely complexity, has been
proposed to measure the cases in which entanglement entropy fails to describe holographically,
such as the growth behaviors of wormhole after the thermal equilibrium.
The concept of complexity in a discrete system is that
the minimum numbers of quantum gates are required to produce
a certain state from a reference state in quantum information theory.

Recently, the complexity definitions from both quantum field theory
and holographic dual viewpoints have been attracted many attentions.
Although many works on the aspect of field theory have been made \cite{Jefferson:2017sdb,
Chapman:2017rqy,Yang:2018nda,Yang:2018tpo,Caputa:2017yrh,Bhattacharyya:2018wym,Khan:2018rzm},
a unique and consistent definition is still missing.
While, from the holographic perspective,
there are two potential prescriptions to realize the complexity,
such as complexity-volume (CV)duality \cite{Stanford:2014jda}
and complexity-action(CA)duality\cite{Brown:2015bva,Brown:2015lvg}.
Since that, a large great of progresses attempting to better understand complexity
from holographic dual point have also appeared in \cite{Carmi:2017jqz,
Mahapatra:2018gig,Lehner:2016vdi,Cai:2017sjv,Cai:2016xho,Alishahiha:2018tep,
Auzzi:2018pbc,Auzzi:2018zdu,Ghodrati:2017roz,Pan:2016ecg,Jiang:2019yzs,
Ghodrati:2018hss,Alishahiha:2017hwg,Jiang:2018pfk,Cano:2018aqi,
An:2018dbz,Jiang:2018sqj,Jiang:2019fpz,Carmi:2016wjl,Kim:2017lrw,
Chapman:2018dem,Chapman:2018lsv,Jiang:2018tlu,Alishahiha:2015rta,
Ghodrati:2019hnn,Ling:2018xpc,Ling:2019ien,Couch:2016exn,Fan:2018wnv,
Fan:2018xwf,Fan:2019mbp,Bernamonti:2019zyy,Caceres:2019pgf,Bernamonti:2020bcf}.

Intriguingly, the investigations on the holographic complexity can be generalized to
states on the dynamical boundary backgrounds, which is associated with different
foliations of the geometry in bulk, such as an asymptotic
Friedman-Lemaitre-Robertson-Walker(FLRW) cosmology boundary \cite{Reynolds:2017lwq},
which might provide us an interesting prototype to understand
the nonperturbative behaviors of cosmology. In Ref.\cite{Apostolopoulos:2008ru,Camilo:2016kxq},
the FLRW Universe on a conformal boundary can be derived from
a class of AdS black holes in bulk and the growth behavior
of holographic complexity of a quantum field on FRLW Universe
is also investigated in \cite{An:2019opz}, where it has been shown that
there are mainly three parts of contributions to the growth rate, namely,
the first one comes from the interaction between a field (or an operator)
on left boundary and one on right boundary, the second one is from the rate of
the spatial volume of the corresponding dual Universe and the third one is from
the constant spatial curvature of the horizon in the AdS-Schwarzschild background.

Motivated by the work in Ref.\cite{An:2019opz}, we would like to extend it to
the cases of some modified gravities, such as Gauss-Bonnet gravity and massive gravity, respectively.
It should be interesting to investigate, for each case, the holographic complexity behavior of
a field state defined on cosmology boundary being dual to a corresponding type of modified gravity,
and we expect that these growth behaviors have some interesting phenomenons.

The paper is organized as follows: in Section II, we simply review the previous work in literature,
where, with the help of the Eddington-Finkelstein coordinates, the FLRW universe metric
can be derived from a large class of static asymptotically $AdS$ black hole spacetimes.
In Section.III, in the Gauss-Bonnet geometry, according to the CV conjecture,
under the large $R_m$ limit, we shall analytically calculate the holographic complexity
growth rate of a field state in the dual universe. In section IV, for a massive gravity case,
we, in a parallel way, investigate the relevant complexity behaviors.
The last section gives the relevant conclusions and discussions.

\section{The metric}
In this section, we briefly review how to derive the FLRW spacetime from AdS black hole
background \cite{Apostolopoulos:2008ru,Camilo:2016kxq}. First of all, a large class of
static asymptotically $AdS_{d+1}$ black hole is described by the metric
\begin{align}\label{met1}
ds^2=-f(r)dt^2+f(r)^{-1}dr^2+\Sigma(r)^2h_{ij}dx^idx^j
\end{align}
where $h_{ij}dx^idx^j$ is line element of the co-dimension two maximally
symmetric subspace with a spherical , planar , or
hyperbolic topology, respectively. The blackening factor $f(r)$
and the function $\Sigma(r)$, which usually can be determined by
Equation of motion in gravity, are required to naturally saturate
the assumption that $f(r)\sim \frac{r^2}{L^2}$ and $\Sigma(r)\sim \frac{r}{L}$
at large $r$ with AdS curvature radius $L$. In order to conveniently
explore holographic complexity behavior of a field living in dual FLRW Universe
embedded into an AdS black hole background,
one can introduce the Eddington-Finkelstein coordinates via
\begin{align}\label{EFCoor}
\nu=t+r^*(r),\ \  dr^*=\frac{dr}{f(r)},
\end{align}
such that one can write the metric (\ref{met1}) in the form
\begin{align}\label{met5}
ds^2=-f(r)d\nu^2+2d\nu dr+\Sigma(r)^2h_{ij}dx^idx^j.
\end{align}
In the following, we shall present a different foliation of
the black hole spacetime (\ref{met5}) in such a way that
the corresponding conformal boundary can take the form of  FLRW spacetime.
For this aim, we need to introduce new time coordinate $V$,\ \ $d\nu=\frac{dV}{a(V)}$,
and the new radial coordinate $R=\frac{r}{a(V)}$, where $a(V)$
explained as the cosmological evolving factor is a positive
function in terms of $V$, in order to reexpress the metric in the form:
\begin{align}\label{met3}
ds^2=2dV dR-[\frac{f}{a^2(V)}-2R\frac{\dot{a}}{a}]dV^2+\Sigma(Ra)^2h_{ij}dx^idx^j.
\end{align}
If taking the large $r$,(or $R$) limit,  then $f(Ra)\sim\frac{(Ra)^2}{L^2}$ and $\Sigma(Ra)\sim\frac{Ra}{L}$ are obtained. Therefore,
the above line element(\ref{met3})in bulk can be approximately replaced by
\begin{align}\label{met4}
ds^2\sim 2dV dR+\frac{R^2}{L^2}[-dV^2+a^2h_{ij}dx^idx^j]
\end{align}
as a result, the new conformal boundary at $R\rightarrow\infty$ has precisely
the desired FLRW universe refereed as the cosmological boundary with spatial curvature $k$.
It's worth noting that it is not the same as the ordinate AdS boundary
at $r\rightarrow\infty$, in which there is a static boundary spacetime.

In the following sections, using the Complexity-Volume conjecture,
we focus on investigating the holographic complexity growth rate of the TFD state
defined on boundary FLRW universe for the AdS black hole background
in the Gauss-Bonnet gravity and massive gravity, respectively.
In particular, we shall analytically compute their time dependence
of complexity and explore the asymptotic growth rate how to relate
a conserved quantity in bulk and the geometrical quantities in the boundary cosmology.
Note that the evolution of dual state depends on two times $t_L$ and $t_R$
denoting the left and right boundary times, respectively.
Without loss of generality, we will adopt the symmetric configuration
times with $t_L=t_R$ as shown in\cite{Carmi:2017jqz}.

\section{The case for the neutral Gauss-Bonnet black hole}
In this section, according to the analysis in literture\cite{An:2019opz},
we use the CV conjecture to explore in detail the behavior of holographic complexity growth rate
with respect to the dual cosmological boundary time in
the Gauss-Bonnet black hole background. The CV conjecture, being related to
the size of an Einstein-Rosen bridge (ERB) to the computational complexity of the dual
quantum field on the boundary, suggests that the complexity is dual to the volume of an extremal codimension-one bulk surface anchored
at the time slice in the boundary on which the state is defined,
\begin{align}
\mathcal{C}_{V}=\frac{max[V]}{G\ell},
\end{align}
where $\ell$ is some additional length scale associated with the bulk geometry, which is usually chosen
to be equal to the effective AdS radius $L_e$ or the AdS radius $L$, as shown in following.
As for the trick, we mainly consult the procedure proposed by \cite{Carmi:2017jqz,An:2018dbz,An:2019opz}.

Let us start by the Einstein-Gauss-Bonnet action consisting of a cosmological constant
and the simplest generalizations of Einstein gravity \cite{Cai:2001dz}, which is
\begin{align}
S=\frac{1}{16\pi G}\int d^{d+1}x\sqrt{-g}[R+\frac{(d-1)(d-2)}{L^2}+\alpha(R_{\mu\nu\gamma\delta}R^{\mu\nu\gamma\delta}-4R_{\mu\nu}R^{\mu\nu}+R^2)]
\end{align}
the metric following from the above action is given in the following form
\begin{align}\label{met2}
ds^2=-f(r)dt^2+f(r)^{-1}dr^2+\frac{r^2}{L^2_{e}}h_{ij}dx^idx^j
\end{align}
where
\begin{align}
f(r)=k+\frac{r^2}{2\tilde{\alpha}}[1-\sqrt{1+4{\tilde{\alpha}}(\frac{\tilde{M}}{r^d}-\frac{1}{L^2})}]
\end{align}
where$\tilde{M}=\frac{16\pi G M}{(d-1)\Omega_{k,d-1}}$ and $\tilde{\alpha}=\alpha(d-2)(d-3)$.
The notation $\alpha$ is the coupling constant of the Gauss-Bonnet term with dimension $(length)^2$ and
the effective AdS radius $L_e$ is obtained by shifting the usual $L$
due to the presence of $\tilde{\alpha}$, namely $L^2_e=\frac{L^2}{2}(1+\sqrt{1-\frac{4\tilde{\alpha}}{L^2}})$.
Note that the effective AdS radius chosen in the metric has some differences from the solution in \cite{Cai:2001dz,An:2018dbz},
but is consistent with the one in \cite{Camilo:2016kxq}.
For the convenience to implement our goals, according to the previous process \cite{Carmi:2017jqz,An:2019opz},
one needs to introduce the Eddington-Finkelstein coordinates, so that the line element (\ref{met2}) becomes
\begin{align}
ds^2=-f(r)d\nu^2+2d\nu dr+\frac{r^2}{L^2_{e}}h_{ij}dx^idx^j.
\end{align}
Here, we have used the effective AdS radius. Next one can early check that the dual boundary cosmology can be derived from
the AdS Gauss-Bonnet black hole, whose result is just the form of Eq.(\ref{met3}).
In order to calculate the time dependence of holographic complexity
in Gauss-Bonnet gravity, we need to embed a surface possessing
the same maximal symmetry as the horizon does,i.e. the surface
is independent of the coordinates $x^i$. As a result, the surface can be described via
the parameterizing equations $\nu=\nu(\lambda)$ and $r=r(\lambda)$ in the parameter $\lambda$.
Then, basing on the method \cite{Carmi:2017jqz,An:2019opz}, its volume is calculated in the following form,
\begin{align}
V&=2\Omega_{k,d-1}L^{d-1}_{e}W\\
W&=\int d\lambda(\frac{r}{L_{e}})^{d-1}\sqrt{-f(r)\nu^{\prime2}+2\nu^{\prime}r^{\prime}}\equiv\int d\lambda\mathcal{L}\label{W}
\end{align}
where the primes indicate the derivatives with respect to $\lambda$.
Since the above integrand $\mathcal{L}$ is not explicitly dependent on $\nu$,
we obtain a conserved quantity $E$ written as
\begin{align}\label{E}
E&=-\frac{\partial\mathcal{L}}{\partial\nu^{\prime}}=(\frac{r}{L_e})^{d-1}\frac{f\nu^{\prime}-r^{\prime}}{\sqrt{-f\nu^{\prime2}+2\nu^{\prime}r^{\prime}}}
\end{align}
we shall refer to it as the energy. Since Eq.(\ref{W})is reparametrization invariant, we
can be free to choose parameter $\lambda$ to keep the radial volume element fixed, namely,
\begin{align}
(\frac{r}{L_{e}})^{d-1}\sqrt{-f(r)\nu^{\prime2}+2\nu^{\prime}r^{\prime}}=1,
\end{align}
such that Equations.(\ref{W})and(\ref{E}) can be rewritten as
\begin{align}
W&=\int^{r_{max}}_{r_{min}}dr(\frac{r}{L_e})^{2(d-1)}\frac{1}{\sqrt{f(\frac{r}{L_e})^{2(d-1)}+E^2}}\\
E&=(\frac{r}{L_{e}})^{2(d-1)}(f\nu^{\prime}-r^{\prime})\label{Energy}.
\end{align}
It is easy to find that Eq.(\ref{Energy}) has an alternative form,
\begin{align}
r^{\prime}&=\sqrt{f(\frac{L_e}{r})^{2(d-1)}+E^2(\frac{L_e}{r})^{4(d-1)}}\label{rprime}
\end{align}
Here, we are assuming a symmetric configuration with $t_L=t_R$ on boundaries,
as a result, the point at $r_{min}$ ,as a minimal radius, should be a turning point of the surface,
and then the derivative $r^{\prime}$ would vanish. Therefore, the minimal radius is determined by
\begin{align}
f(r_{min})(\frac{r_{min}}{L_e})^{2(d-1)}+E^2=0.
\end{align}
As shown in literature\cite{Carmi:2017jqz}, the turning point is behind the horizon and hence
we have $f(r_{min})<0, r^{\prime}=0$ and $\nu^{\prime}>0$. Thus,
we can deduce a useful result that $E<0$ by calculating Eq.(\ref{Energy}) at the turning point.
Making the use of Eqs.(\ref{Energy}) and (\ref{rprime}), one has
\begin{align}
t_R+r^*(r_{max})-r^*(r_{min})=\int^{r_{max}}_{r_{min}}dr[\frac{E}{f\sqrt{f(\frac{r}{L_e})^{2(d-1)}+E^2}}+\frac{1}{f}].
\end{align}
Here, we have used a fact that, at the innermost point or turning point, there is $t=0$ due to the symmetry.
Meanwhile, the above equation both sides multiply with the conserved quantity $E$, we obtain
\begin{align}
W=\int^{r_{max}}_{r_{min}}dr\frac{\sqrt{f(\frac{r}{L_e})^{2(d-1)}+E^2}+E}{f}-E[t_R+r^*(r_{max})-r^*(r_{min})].
\end{align}
Now, we turn to considering the growth behavior for time dependent holographic complexity
in the Gauss-Bonnet theory. Hence, adopting the new time coordinate $V$
and radial coordinate $R$, and employing the chain rule of differentiation,
 \begin{align}
\frac{\partial W(V_R,R_{max})}{\partial V_R}=\frac{\partial W}{\partial t_R}\frac{\partial t_R}{\partial V_R}
+\frac{\partial W}{\partial r_{max}}\frac{\partial r_{max}}{\partial V_R},
\end{align}
the partial derivatives can be given as
\begin{align}
\frac{\partial W}{\partial t_R}&=-E\\
\frac{\partial t_R}{\partial V_R}&=\frac{1}{a(V_R)}-\frac{R_{max}\dot{a}(V_R)}{f(R_{max}a)}\\
\frac{\partial W}{\partial r_{max}}&=\frac{\sqrt{f(r_{max})(\frac{r_{max}}{L_e})^{2(d-1)}+E^2}}{f(r_{max})}\\
\frac{\partial r_{max}}{\partial V_R}&=R_{max}\dot{a}(V_R)
\end{align}
where the dots denote the derivatives with respect to $V$. Therefore, utilizing above results, we arrive at
\begin{align}
\frac{\partial W(V_R,R_{max})}{\partial V_R}=-E[\frac{1}{a(V_R)}-\frac{R_{max}\dot{a}(V_R)}{f(R_{max}a)}]
+\frac{\sqrt{f(R_{max}a(V_R))(\frac{R_{max}a(V_R)}{L_e})^{2(d-1)}+E^2}}{f(R_{max}a(V_R))}R_{max}\dot{a}(V_R).
\end{align}
So far, the analysis is quite general. From now on, we focus on the growth rate of holographic complexity
in $d+1=5$ dimensional Gauss-Bonnet gravity. Taking the limit $R_{max}\rightarrow\infty$, and keeping $V_R$ fixed,
we come to the conclusion that the growth rate with respect to the cosmology boundary time is given by
 \begin{align}\label{result1}
\frac{\partial\mathcal{C}}{\partial V_R}&=\frac{1}{GL_e}\frac{\partial V}{\partial V_R}\\\nonumber
&=-\frac{2\Omega_{k,3}L^2_e}{G}\frac{E}{a(V_R)}+\frac{2\Omega_{k,3}}{G}[\frac{R^3_{max}}{3}\frac{d(a^3(V_R))}{dV_R}-\frac{A}{2}R_{max}\dot{a}(V_R)+\ldots]
\end{align}
where $A=\frac{2\tilde{\alpha}k}{1-\sqrt{1-\frac{4\tilde{\alpha}}{L^2}}}$. Here we have used $\ell=L_e$.
To summarize, taking the large $R_{max}$ condition and keeping the time $V_R$ fixed,
for holographic complexity of a field on the FLRW universe
being dual to a five dimensional AdS Gauss-Bonnet black hole, we come to
a conclusion that its growth behavior includes the first term, as a finite term,
mainly relating to the conserved charge $E$ and the second term
(the leading divergent term) being proportional to the rate of the spatial volume
in the dual FLRW universe as well as the third term (the sub-leading divergent term)
coming from the codimension-two spatial constant curvature, whose behaviors are very like
the one presented in literature \cite{An:2019opz}. It has been shown that,
according to the above evaluating result(\ref{result1}),
the Gauss-Bonnet effect plays an interesting role on the such a complexity growth.
When the parameter $\tilde{\alpha}\rightarrow 0$, its growth rate in AdS-Gauss-Bonnet black hole
reduces to the result in AdS-Schwarzschild case.

\section{The case for the neutral massive black hole}
In the section, we are going to study the counterpart
in massive black hole in a parallel way. First, let us review briefly
the solution from massive gravity in five dimensional spacetime. And then,
we shall demonstrate how the growth behavior of holographic complexity
of a quantum field in the FLRW universe link the conserved charge
and some geometric quantities on boundary cosmology by holographic dual,
and what role the graviton mass term can play on this growth rate.
First, let us write down the neutral massive Einstein action
 consisting of the Ricci scalar, cosmological constant term,
 graviton mass terms\cite{deRham:2010kj,Vegh:2013sk,Cao:2015cti,Cai:2014znn,Cao:2015cza},
 which can be expressed as
\begin{align}\label{S}
S=\int d^5x\sqrt{-g}[\frac{1}{2\kappa^2}(R-2\Lambda)+\frac{m^2}{2\kappa^2}(c_1u_1+c_2u_2+c_3u_3+c_4u_4)],
\end{align}
where
\begin{eqnarray}
u_1&=&tr\mathcal{K},\\\nonumber
u_2&=&(tr\mathcal{K})^2-tr(\mathcal{K}^2),\\\nonumber
u_3&=&(tr\mathcal{K})^3-3tr\mathcal{K}tr(\mathcal{K})^2+2tr(\mathcal{K}^3),\\\nonumber
u_4&=&(tr\mathcal{K})^4-6tr(\mathcal{K}^2)(tr\mathcal{K})^2+8tr(\mathcal{K}^3)tr\mathcal{K}+3(tr(\mathcal{K}^2))^2-6tr(\mathcal{K}^4)
\end{eqnarray}
$c_1$, $c_2$ ,$c_3$ and $c_4$ are negative constants, but $c_0$ is a positive constant; $\kappa^2=8\pi G$, and
the matric $\mathcal{{K^{\mu}}_{\nu}}$ is defined by
$\mathcal{{K^{\mu}}_{\nu}}=\sqrt{g^{\mu\alpha}f_{\alpha\nu}}$.
It tells us that the graviton is allowed to obtain its mass $m$
\footnote{Here, If requiring the graviton mass $m^2\in (0,\frac{12c_2-12c_2\sqrt{1+\frac{L^2c^2_1}{6c^2_0c^2_2}}}{L^2c^2_1})$, then the single horizon appears, as a consequence, it shares the similar Penrose diagram with the neutral AdS-black hole.} by the reference
metric coupling the bulk metric to break differemophism symmetry.
Here following the ansatz in \cite{Cai:2014znn}, the reference metric without dynamical behavior
is chosen as $f_{\mu\nu}=diag(0,0,c^2_0h_{ij})$.
Its solution following from the above action was given in static coordinate as
\begin{align}\label{Solu}
ds^2=-f(r)dt^2+\frac{dr^2}{f}+\frac{r^2}{L^2}h_{ij}dx^idx^j
\end{align}
where
\begin{align}
f(r)=k+\frac{r^2}{L^2}-\frac{m_0}{r^2}+\frac{c_0c_1m^2}{3}r+c^2_0c_2m^2+\frac{2c^3_0c_3m^2}{r}+\frac{2c^4_0c_4m^2}{r^2}
\end{align}
where $m_0$ is related to the mass parameter of black hole in massive gravity in five dimensional geometric configuration, namely
\begin{align}
M=\frac{3\Omega_3m_0}{16\pi G}
\end{align}
similarly, when taking the large $r_{max}=R_{max}a$ limit,
the solution (\ref{Solu}) can give rise to the form of metric (\ref{met4}).
Next, using the symmetric configuration times with $t_L=t_R$ as stated above
and the corresponding metric form(\ref{met4}),
we shall concentrate on time dependent complexity in five dimensional manifold and evaluate
the volume of Einstein-Rosen Brige at a specific boundary time, its
mathematical relation is similarly given by
\begin{align}
V&=2\Omega_{k,3}L^3W\\
W&=\int^{\lambda_{max}}_{\lambda_{min}}(\frac{r}{L})^3\sqrt{-f(r)\nu^{\prime2}+2\nu^{\prime}r^{\prime}}d\lambda
\end{align}
Repeating the work in a similar way, it is not hard to find that
\begin{align}\label{W3}
W=\int^{r_{max}}_{r_{min}}dr\frac{\sqrt{(\frac{r}{L})^{6}f(r)+E^2}+E}{f}-E[t_R+r^*(r_{max})-r^*(r_{min})].
\end{align}
Here, the physical meaning of $E$ is also a conserved charge on the ERB in such system.
Applying the chain rule of differentiation and the new coordinates above, we can easily find
 \begin{align}
\frac{\partial W(V_R,R_{max})}{\partial V_R}=\frac{\partial W}{\partial t_R}\frac{\partial t_R}{\partial V_R}
+\frac{\partial W}{\partial r_{max}}\frac{\partial r_{max}}{\partial V_R}.
\end{align}
The partial derivatives can be exactly given as
\begin{align}
\frac{\partial W}{\partial t_R}&=-E\\
\frac{\partial t_R}{\partial V_R}&=\frac{1}{a(V_R)}-\frac{R_{max}\dot{a}(V_R)}{f(R_{max}a)}\\
\frac{\partial W}{\partial r_{max}}&=\frac{\sqrt{f(r_{max})(\frac{r_{max}}{L})^{6}+E^2}}{f(r_{max})}\\
\frac{\partial r_{max}}{\partial V_R}&=R_{max}\dot{a}(V_R).
\end{align}
Hence, plugging the above partial derivatives relations into Eq.(\ref{W3}), one naturally arrives at
\begin{align}
\frac{\partial W(V_R,R_{max})}{\partial V_R}=-E[\frac{1}{a(V_R)}-\frac{R_{max}\dot{a}(V_R)}{f(R_{max}a)}]
+\frac{\sqrt{f(R_{max}a(V_R))(\frac{R_{max}a(V_R)}{L})^{6}+E^2}}{f(R_{max}a(V_R))}R_{max}\dot{a}(V_R).
\end{align}
From the above equation, after taking the large $R_{max}\equiv R_m$ expansion
and keeping the time $V_R$ fixed, it can approximately reduce to
\begin{align}\label{result2}
\frac{\partial C_V}{\partial V_R}=\frac{2\Omega_{k,3}}{G}[\frac{-EL^2+\dot{a}(V_R)(\frac{3}{4}AB-\frac{C}{2}-\frac{5}{16}A^3)}{a(V_R)}+\frac{1}{3}R^3_m\frac{da^3}{dV_R}
-\frac{A}{4}R^2_m\frac{da^2}{dV_R}+R_m\dot{a}(3A^2-\frac{B}{2})+\ldots]
\end{align}
where the constants in the above result are $A=\frac{c_0c_1m^2L^2}{3}$, $B=L^2(k+c^2_0c_2m^2)$,
and $C=2c^3_0c_3m^2L^2$,respectively. We have used $\ell=L$ for this case. So far, under the same conditions,
we have captured the holographic complexity growth behaviors for the TFD state
on a dual FLRW universe in massive gravity. The first divergent term being proportional to
the growth of the spatial volume of the Universe on the boundary is still given, which is in agreement with
the relevant result from the last model. In contrast with the relevant results
in the previous model or the ones in \cite{An:2019opz}, we find that there are some new
interesting phenomenons in the other terms due to the graviton mass effect.
The new intriguing phenomenons are that the finite term consists of,
beside the usual conserved charge $E$, the novel contribution from graviton mass effect;
and the third divergent term also presents both the spatial curvature of the horizon
$k$ and graviton mass effect; furthermore, to be more interesting, the second divergent
term is totally caused by the rate of the area of the dual Universe
on account of the graviton mass effect. This ,In this sense, means that
the graviton mass effect plays a vital role on the growth behaviors of
a conformal field state defined on the dual FLRW Universe.

\section{conclusions and discussions}
In this paper, inspired by the recent work\cite{An:2019opz},
using the complexity-volume duality and the prescription on the time dependence of
holographic complexity \cite{Carmi:2017jqz}, we have holographically computed the growth behaviors of complexity
for a field in the boundary cosmology being dual to Gauss-Bonnet gravity and massive gravity in bulk, respectively.

In the framework context of Gauss-Bonnet gravity, under the large $R_m$ condition,
we have found a conclusion that the change behavior of holographic complexity
for a field defined in the dual Universe located at an asymptotic AdS boundary
is mainly governed by a finite term relating to the conserved charge $E$ and the leading divergent term
being proportional to the rate of the spatial volume in the dual FLRW universe
as well as the sub-leading divergent term coming from the contribution of
the codimension-two spatial constant curvature, whose behaviors are very like
the one presented in literature \cite{An:2019opz}. We also note that,
according to the result of Eq.(\ref{result1}), the Gauss-Bonnet effect
specifically plays an interesting role on the such a complexity growth,
which explicitly makes our result distinguish from the one from the standard Einstein gravity case.

Similarly, in contrast to the relevant result in the Gauss-Bonnet gravity
or the one in \cite{An:2019opz}, for the massive gravity case,
except the first divergent term obeying the growth rate of the spatial volume of
the Universe located at an asymptotic AdS-boundary, there are some new remarkable results to be observed
in the other terms due to the graviton mass effect. We have demonstrated that,
under the same conditions, the new intriguing phenomenons are that, beside
the conserved charge $E$, the some graviton mass effect also contributes the finite term
; and the third divergent term is determined by the spatial curvature of the horizon
$k$ and graviton mass effect; furthermore, to be more interesting and surprising,
the graviton mass effect can be completely responsible for the second divergent
term as an new additional term saturating an area law. Thus, they allow
us to distinguish clearly from the results from the AdS-Schwarzschild case.

According to the information of the above results, our models have explicitly exhibited
some universal complexity growth behaviors. Firstly, the finite term is directly proportional
to some conserved quantities and inversely proportional to the cosmological factor,
which can be interpreted as a interacting term between two localized operators
at the left and right boundaries \cite{An:2019opz}, respectively. Secondly, the leading divergent term
in each case obeys a volume law, which at the qualitative level, is quite consistent
with the definition of complexity from field theory landscape. Lastly,
the effects from the Gauss-Bonnet term or the graviton mass terms give their contributions to
govern the evolution of complexity in such holographic dual.
It may also have implicitly suggested that the nonperturbative evolution
properties of a field in the FLRW universe can be implemented
in such holographic dual,as mentioned in literatures \cite{Camilo:2016kxq,An:2019opz}.

Furthermore, applying the CA conjuncture in the holographic context, one will attempt to explore such data of
holographic complexity in the above each case and desire to capture the similar results in future.

\begin{acknowledgments}
  Wen-Jian Pan would like to thank Shan-Ming Ruan, Peng Liu and Min-Yong Guo
 for useful discussions. This work is supported by the National Natural Science
 Foundation of China (Grant Nos. 11847121 and 11704210.)
\end{acknowledgments}

\end{document}